\documentclass[sigconf]{acmart}
\usepackage[normalem]{ulem}
\usepackage{xspace}
\usepackage{tabularx}
\usepackage[keeplastbox]{flushend}

\newcommand{\one}{({\em i}\/)\xspace}
\newcommand{\two}{({\em ii}\/)\xspace}

\newcommand{\tool}{{\sc Pythia}\xspace} 

\copyrightyear{2019}
\acmYear{2019} 
\setcopyright{iw3c2w3}
\acmConference[WWW '19]{Proceedings of the 2019 World Wide Web Conference}{May 13--17, 2019}{San Francisco, CA, USA}
\acmBooktitle{Proceedings of the 2019 World Wide Web Conference (WWW '19), May 13--17, 2019, San Francisco, CA, USA}
\acmPrice{}
\acmDOI{10.1145/3308558.3313664}
\acmISBN{978-1-4503-6674-8/19/05}

\begin{document}

\date{}

\title[\tool]{\tool: a Framework for the Automated Analysis of Web Hosting Environments}

\author{Srdjan Matic}
\affiliation{
 \institution{University College London}
}
\email{s.matic@cs.ucl.ac.uk}

\author{Gareth Tyson}
\affiliation{
 \institution{Queen Mary University of London}
}
\email{gareth.tyson@qmul.ac.uk}

\author{Gianluca Stringhini}
\affiliation{
 \institution{Boston University}
}
\email{gian@bu.edu}

\begin{abstract}
  \noindent
  A common approach when setting up a website is to utilize third
  party Web hosting and content delivery networks. Without taking this
  trend into account, any measurement study inspecting the deployment
  and operation of websites can be heavily skewed. Unfortunately, the
  research community lacks generalizable tools that can be used to
  identify \emph{how} and \emph{where} a given website is
  hosted. Instead, a number of ad hoc techniques have emerged, e.g.,
  using Autonomous System databases, domain prefixes for CNAME
  records.  In this work we propose \tool, a novel lightweight
  approach for identifying Web content hosted on third-party
  infrastructures, including both traditional Web hosts and content
  delivery networks. Our framework identifies the organization to
  which a given Web page belongs, and it detects which Web servers are
  self-hosted and which ones leverage third-party services to provide
  contents. To test our framework we run it on 40,000 URLs and
  evaluate its accuracy, both by comparing the results with similar
  services and with a manually validated groundtruth. Our tool
  achieves an accuracy of 90\% and detects that under 11\% of popular
  domains are self-hosted. We publicly release our tool to allow other
  researchers to reproduce our findings, and to apply it to their own
  studies.
\end{abstract}

\ccsdesc[300]{Networks~Public Internet}

\keywords{CDNs, infrastructure, topology, web}

\maketitle


\pagenumbering{arabic}

\section{Introduction}
\label{sec:intro}
When deploying a website, companies have the choice to either host it
on their own servers, or to offload the content to third-parties,
e.g., Web hosts or Content Delivery Networks (CDNs). Choosing a third
party can have multiple advantages, including cost savings,
reliability, and the ability to sustain larger amounts of traffic
(even during distributed denial of service attacks).  We argue that
understanding the Web hosting landscape is important for a number of
reasons.  These range from allowing us to assess how critical certain
hosting infrastructures are and to estimate the impact that a network
attack could have on the
Web~\cite{delignat2015network,liang2014https,simenovski2017who}, to
being able to determine who is responsible when incidents (e.g.,
malware hosting) occur~\cite{tajalizadehkhoob2017role}.  Despite the
importance of the problem, the research community lacks scalable
methods to map the hosting landscape.  Instead, multiple studies tend
to take an ad hoc approach, relying on various assumptions, which
differ between papers. Although there are third party services that
offer this functionality~\cite{netcraft,webhosting}, they do not
disclose their methodology, creating concerns for both
reproducibility, as well as accuracy of their results.

To fill this gap, this paper \textit{presents an open source tool to
  the community,\footnote{The source code of \tool is available at
    \url{https://bitbucket.org/srdjanmatic/pythia.git}} which can
  determine whether the webpage of an organization is self-hosted or
  it uses a third-party hosting provider}. Our tool, \tool, leverages
the HTML code of the webpage, domain information, and the network
ownership obtained from RDAP records, to determine if the content is
hosted on a third-party infrastructure. This is done by computing the
ownership of both the webpage \emph{and} the hosting provider, such
that the two can be compared. \tool is built with a modular design and
it is capable of obtaining information of the landing webpage even in
the presence of complex HTML structures which use redirects.
  
To evaluate the efficacy of \tool, we run it on 40,000 URLs generated
from the Alexa top-10k domains~\cite{alexatop1m}. Our validation
process shows that our framework outperforms similar applications
available on the Web, and it achieves an accuracy of 90\% in detecting
when a webpage is hosted by a third party. Furthermore, our
measurement reports that over 89\% of the popular domains that we
inspected, take advantage of third parties for their hosting needs.

\tool is open source and allows the research community to reproduce
our findings.  We intend this to become a shared community effort,
allowing third party researchers to avoid the complexity involved in
devising and building their own independent methodologies for this
commonly encountered task.


\section{Background}
\label{sec:overview}
Understanding and measuring the Web hosting ecosystem is a complex
endeavor. To complete this task, we need both information about the
ownership of domains and the ownership of the IP addresses where
webpages are hosted. In this section we introduce the concepts on
which our approach is based, and the type of data that we retrieve to
determine whether webpages are self-hosted or not.

\subsection{Third-Party Hosting}
Content Delivery Networks (CDNs) and Hosting Services are two popular
mechanisms for delivering content to end users on behalf of other
organizations. By offloading the task of serving the content of a
website to third parties, these solutions are designed to provide
better availability, scalability, faster content loads, redundancy,
and enhanced security. These technologies have become so widespread
that according to recent statistics, more than 60\% of the most
visited websites use CDNs to serve content to their
users~\cite{cdnusage}. In this work we study the deployment of any
kind of solution that delivers Web content for third parties, and for
this reason we use the term \textit{hosting} to refer to the
\textit{network where Web servers offering a service are based}.

There are countless papers that have explored the hosting patterns of
websites, each taking a slightly different approach.  A common
approach is to launch large-scale distributed
measurements~\cite{su2009drafting,ager2011web,fanou2016}, which
perform DNS queries around the world to retrieve and classify DNS
responses. This, unfortunately, is extremely complex and costly;
furthermore, it cannot alone confirm if the infrastructure is
third-party operated without further inspection. Calder et
al.~\cite{calder2013mapping} utilized the EDNS-0 Client Subnet
extension to simulate distributed queries towards Google's
CDN. Although it revealed a large number of servers, all were operated
by Google rather than third-parties. These techniques also do not work
well for Anycast CDNs~\cite{calder2015analyzing}, which do not
necessarily return DNS responses containing redirects. Another
strategy employed is to utilize domain prefix lists, which map CNAME
responses to their respective CDNs~\cite{scheitle2018}. These,
however, are limited to CDNs that exclusively rely on CNAME redirects
(e.g., this excludes Bing).  Furthermore, the list requires constant
maintenance to remain up-to-date. Lastly, some studies utilize IP
address to Autonomous System (AS) mappings~\cite{ibosiola2018movie} or
metadata encoded into DNS records~\cite{Bottger18}; these, however are
vulnerable to misattributing ownership, e.g, when a CDN places a cache
in a third-party network. Such techniques have also been complimented
with manually curated AS annotations, which stipulate the type of
AS~\cite{noroozian2016gets}. Again, these suffer from both manual
annotation errors and require substantial upkeep.  We argue that these
diverse ad hoc techniques are driven by the lack of a standardized
tool within the community, which can provide metadata on website
hosting patterns.

\subsection{The RDAP Protocol}
To acquire information about a domain's ownership we use the
Registration Data Access Protocol (RDAP)~\cite{RFC7482}. This protocol
was designed to replace the WHOIS~\cite{RFC3912} protocol as the
authoritative source for registering information about IP addresses,
ASes and domain names. While its predecessor retrieved free text
content, RDAP leverages a RESTful interface to deliver the data in a
machine-readable JSON format. This simplifies the parsing process and
allows us to easily extract information (e.g., the type of entity to
which a range of IP addresses has been assigned, or the description of
an AS). In this paper, we use the RDAP protocol to retrieve
information about the ownership of an IP address to which a domain
resolved.


\section{Overview of \tool}
\label{sec:approach}

\begin{figure}[t]
  \centering \includegraphics[scale=.3]{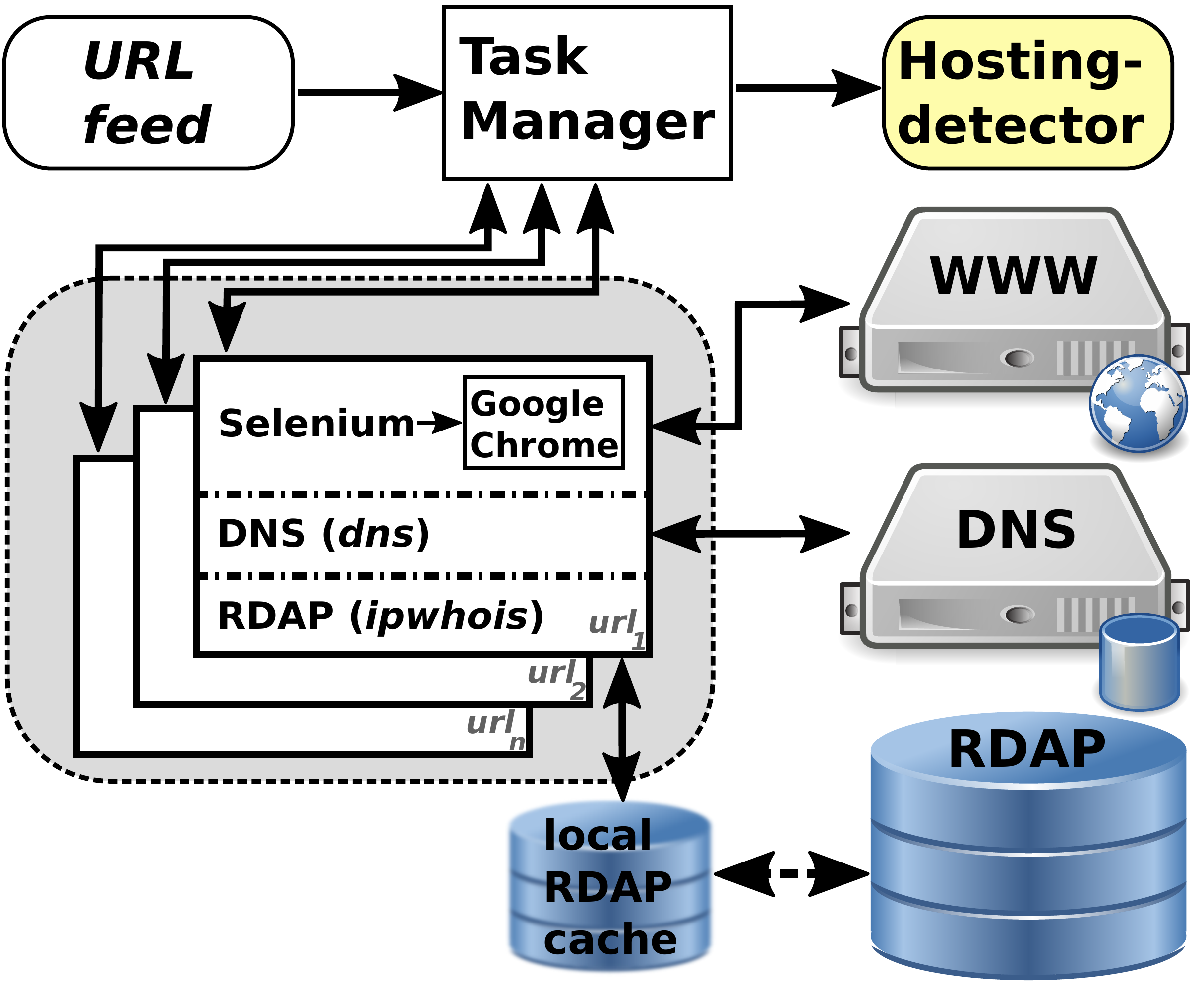}
  \caption{An overview of \tool. The task manager processes URLs from
    a queue and instantiates a \textit{crawler} task for each URL. The
    crawler collects information about a webpage and the environment
    from which the page was retrieved. The output of the crawler is
    stored in a dedicated data structure, which is later analyzed with
    the module that checks if content was served from a self-hosted
    network environment.}
  \label{fig:tool_overview}
\end{figure}

\tool is entirely written in Python, and
Figure~\ref{fig:tool_overview} provides an overview of its system
components.  In a pipeline, \tool performs the following two steps,
\one~\emph{Data Collection:} it takes a URL feed, and renders the list
of websites, recording detailed information on all resources loaded,
before complementing the data with RDAP records for the IP address
hosting the website; and \two~\emph{Hosting Detector:} it then passes
this information through a hosting detector, which decides if the
owner of the webpage is the same as the owner of the hosting
infrastructure where the server is located.  The outcome is a
structured JSON file that details if the website is self-hosted or
operated by a third-party. In addition to this, \tool also provides
information on the ownership of both the webpage and the hosting
service.

\subsection{Data Collection}
We first present the methods we use to collect the necessary data to
infer ownership.  This includes Web data, DNS information and,
finally, RDAP records for all domains loaded.

\paragraph{Web Data Collection}
Upon receiving a list of URLs, the crawler obtains Web content through
Selenium,\footnote{\url{https://www.seleniumhq.org/}} a popular
framework used for testing Web applications. We instrument Selenium to
take a URL and to render it within a fully fledged instance of the
Google Chrome browser. After the page has been loaded and rendered,
our module outputs the retrieved HTML, in addition to a list of all
the HTTP requests/responses (URLs) that were generated during the
process. Each request/response is accompanied by metadata including
the HTTP status code, and HTTP headers (e.g., server, content-type).

An important part of the crawling process is to determine when a
webpage has finished loading. Since we do not know how long this
process will take, we use an adaptive mechanism that leverages the
information logged by the Web browser. We continuously monitor the
browser logs and we consider the page loaded once all the requests to
external resources have received their corresponding response. At the
same time, we also set a hard timeout, after which we will close the
browser session independently if the loading succeeded or not.

As part of the process, we also follow all the redirects that occur
while loading an URL. This includes not just CNAME or HTTP redirects,
but also those triggered by the refresh meta tag or a script. However,
being primarily a tool for developers who need to check their
own webpages, Selenium does not provide access to the browser
internals. Hence, \tool extracts information about the redirection
chain from the browser log. After filtering out all the request to
external resources, we identify the landing page and URLs on which the
browser terminated the navigation.

\paragraph{Domain Resolution} Since Web browser logs do not contain
information about domain resolution, \tool launches DNS queries for
all domains encountered (at a modest cost of only around 3.2\%
additional overhead per webpage). An advantage of this systems is that
we can to use our own DNS server and we do not need to rely on the
built-in mechanisms of proprietary resolvers, which might be hardcoded
in the browser.

\paragraph{IP Ownership} Using the DNS results, we then 
determine the ``owner'' of the IP space where the content is hosted.
\tool uses the RDAP protocol to find the network prefix to which an IP
address belongs to, and to identify the owner of that range. Our
framework uses a local RDAP cache to overcome rate-limiting issues of
RDAP servers and to avoid querying ranges for which we already have
fresh information. After successfully completing the RDAP resolutions,
all the data generated by the above steps is stored in a JSON data
structure.

\subsection{Detecting Hosting Infrastructures}
The next step is to use the above information to detect if a website
is self-hosted, or whether it uses on a third-party
infrastructure. Our tool identifies the organization/company behind a
domain name and a webpage, and searches for evidence that the page
owner is the same as the owner of the network prefix or the AS hosting
the Web server. We do not differentiate among various types of hosting
services (i.e., VPS, CDN, or generic web hosting) and we do not use
any precompiled list of popular or known hosting services. Instead we
extract our information from the URL and the HTML retrieved from the
landing page, and we match this data with the RDAP response of the IP
address that is hosting the web server.  This process allows us to
detect the third-party network infrastructures even in the presence of
CDN caches located at ISPs: even if we did not identify correctly the
provider, our algorithm will detect a mismatch in the ownership of the
webpage and the IP range.

To identify the organization that owns a webpage we use both the
information from the URL and the HTML code. In particular, from the
URL component we extract the Effective Second Level Domain (ESLD), and
from the HTML we use the content of the $<$title$>$ tag. Before
retrieving any ownership information for a RDAP response, we first
filter unnecessary details from the data such as ``comments'' or the
``symbolic name of the network'', which can contain references to the
owner of the webpage even when the IP range is assigned to a
completely different organization. After this step, each string
contained in the \textit{HTML title} or the \textit{RDAP fields} has
its leading space delimiters removed, is cleaned from punctuation
characters and stop words, converted into lower case, and finally is
split into tokens on space delimiters. The DNS system does not allow
domain names to contain space delimiters and it is common have
domains, such as ``bankofamerica.com'', where the ESLD is a
combination of multiple words. To overcome this issue, the ESLD string
follows the same cleaning process of the title and the RDAP, with the
only difference in the tokenization, which is performed following the
technique described in~\cite{segaran2009beautiful}.

This process results in a series of string tokens that represent ownership 
features of both the webpage and the domain/IP address hosting it. 
The next step is to compare these tokens to see if they correspond. 
Our algorithm does six checks: four with the strings contained in the
title/ESLD/RDAP, and an additional two with the tokenized versions of
those strings. First, the algorithm verifies if the HTML title or the
ESLD appears as a sub-string in any of the RDAP
fields. Subsequently it repeats the same procedure with each string in
the RDAP fields by comparing it both with the HTML title and the
ESLD. The output of this process is binary and if the algorithm
finds a match, it concludes that the owner of the webpage is also the
owner of the network. As a final step, the algorithm checks for the
presence of common tokens among the lists of tokens obtained from the
HTML title/ESLD and the RDAP information. In this case a single match
is not enough to conclude that the same organization owns both the
webpage and the network, and we require that the common tokens
represent at least 50\% of the overall number of tokens in the
shortest list.


\section{Validation and Evaluation}
\label{sec:evaluation}
\begin{table*}[t]
  \tiny
  \centering
  \begin{tabularx}{\textwidth}{|X|X|X|X|X|X|X|X|X|X|X|X|X|}
    \cline{2-13}
    \multicolumn{1}{c|}{} & \multicolumn{9}{c|}{\sc URLs (Domains)} & \multicolumn{3}{c|}{\sc IPs}\\
    \cline{2-13}
    \multicolumn{1}{c|}{} & {All} & {\sc Starting} & \multicolumn{4}{c|}{\sc Crawls} & \multicolumn{3}{c|}{\sc Landing} & \multicolumn{3}{c|}{}\\
    \hline
    {Format}
    & {\sc Tot.} & {\sc Tot.} & {\sc Completed} & {\sc Domain Change} & {\sc Protocol Change} & {\sc AVG. Redirects} & {\sc NON-triggering exceptions} & {\sc self-host} & {\sc 3rd-host}
    & {\sc Tot.} & {\sc NON-triggering exceptions} & {\sc Landing Domains}\\
    \hline
    {http} & 874,574 (46,679) & 10,000 (10,000) & 9,204 & 6,114 & 6,701 & 2.6 & 8,968 (8,897) & 941 (935) & 8,027 (7,962) & 34,332 & 33,728 & 8,178\\
    \hline
    {http + \newline www} & 885,035 (43,359) & 10,000 (10,000) & 9,280 & 2,606 & 6,627 & 2.4 & 9,033 (8,962) & 977 (971) & 8,056 (7,991) & 32,496 & 31,973 & 8,214\\
    \hline
    {https} & 745,939 (40,696) & 10,000 (10,000) & 8,215 & 5,147 & 580 & 2.3 & 8,037 (7,989) & 877 (873) & 7,160 (7,116) & 30,640 & 30,217 & 7,353\\
    \hline
    {https + \newline www} & 794,449 (39,787) & 10,000 (10,000) & 8,673 & 2,229 & 491 & 2.2 & 8,462 (8,404) & 918 (913) & 7,544 (7,491) & 30,177 & 29,781 & 7,683\\
    \hline
    \hline
    {All} & 1,736,929 (54,410) & 40,000 (20,000) & 35,372 & 16,096 & 14,399 & 2.4 & 13,940 (11,253) & 1,559 (1,220) & 12,381 (10,033) & 38,092 & 37,492 & 9,188\\
    \hline
  \end{tabularx}
  \caption{Results of running \tool on the \textsc{top-40k-URLs}
    dataset. The values in the brackets indicates the unique number of
    elements for each entry.}
  \label{tbl:data_collection}
\end{table*}

\tool is intended to be both accurate and straightforward to use for
the community.  To validate its capabilities we next run it over
multiple datasets.

\subsection{Validating the Data Collection}
Our first goal is to test the efficacy of our crawler in collecting
the necessary data to perform the hosting classification.  Hence, we
run \tool over a series of URL lists.  A first dataset includes 10,000
unique domain names obtained from a snapshot of the ``Alexa top 10,000
websites'' (\textsc{top-10k}) on 1st of May 2018. A second dataset
(\textsc{top-20k-www}) is an extended version of the previous one,
where domains are extended with the ``www.''  prefix. Finally, a third
dataset (\textsc{top-40k-URLs}) includes all the entries from
\textsc{top-20k-www} expanded with ``http://'' and ``https://''
prefixes.

\paragraph{Data Collection}
We run \tool over the \textsc{top-40k-URLs} to collect information
about their home pages. 
We split our dataset into chunks of 350 elements and we process each
chunk separately. Each element of the chunk is a unique domain name,
which is crawled both with and without the ``www.'' prefix and with
the two protocols that \tool supports (HTTP and HTTPS). This means
that when we successfully crawl an entire chunk, we obtain information
for 1400 unique URLs. We refer to the initial URL from which we begin
our crawling, and that we load in the browser, as ``starting URL'';
similarly, the URL on which the crawl terminates is called ``landing
URL''. Once a chunk is processed, \tool waits for 120 seconds before
switching to the next one. Each chunk is analyzed using 20 parallel
instances of our Crawler module, which uses a maximum timeout of 60
seconds while waiting for a webpage to finish loading. Note that when
collecting IP Ownership information from RDAP, we randomize waiting
timeouts, with of a maximum of 90 seconds, before retrying a query
that triggered an exception; after 3 consecutive exceptions, the
module marks an IP as ``no info available'' before switching to the
next one.  The entire data collection process took place from a single
machine, although we note it is possible to split the dataset and run
parallel instances of \tool on different
machines.

\begin{figure}[t]
  \centering
  \includegraphics[scale=.32]{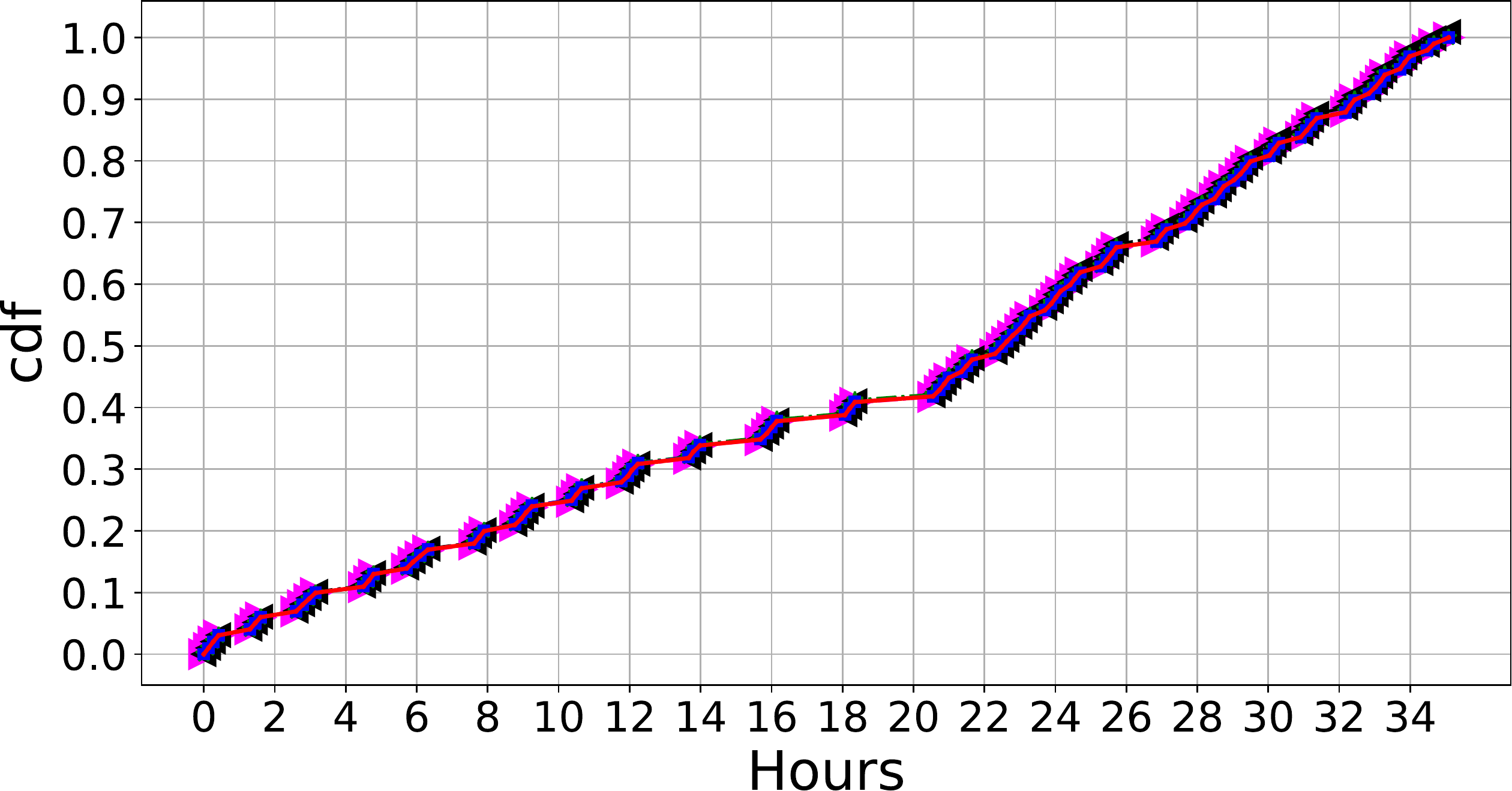}
  \caption{Cumulative distribution functions of the successful crawls
    with different URL formats.}
  \label{fig:cdf_successuful_crawls}
\end{figure}

\paragraph{Data Collection Performance}
The overall process of downloading the HTML, resolving DNS names and
collecting RDAP data, took 35 hours to complete for the
\textsc{top-40k-URLs}.  Figure~\ref{fig:cdf_successuful_crawls}
presents the Cumulative Distribution Function (CDF) of the increase of
number successful crawls across time. To be able to use fresh entries
from our local RDAP cache, we crawled the 4 starting URLs linked to
each domain at the same time. Due to this choice, the CDFs of each
``URL format'' have very similar shapes. Hence \textit{four}
distributions in Figure~\ref{fig:cdf_successuful_crawls} are almost
stacked on the top of each other, and the red line connecting the
values is the average across those distributions. After 24 hours, we
had crawled only 50\% of the URLs, and in the last 1/3 of the time, we
obtained the information for the remaining half of the dataset. The
dataset contained shuffled entries, which did not follow any ranking
depending on the ``popularity of a domain'' and the likelihood to have
an ``unreachable URL'' at the beginning or at the end the crawl is the
same. We explain the spike in the increase of the number of downloads
after the 24th hour with the presence of our RDAP cache. As we will
later show, the majority of URLs/domains use a third-party hosting
provider and as times passes we observe an increase of the number of
RDAP queries which can be resolved with our local cache. Those local
resolutions increase our crawling speed allowing us to allow us to
gather information for the same amount of elements, in half of the
time.

In Table~\ref{tbl:data_collection} we summarize the results of the
data collection process using \tool. The first thing to notice is that
85\% of the starting URLs are successfully reached. Overall, our
crawler visited 1,736,929 external URLs, which were retrieved from
54,410 different domains. This suggests an average factor of 42 URLs
per ``starting URL''. It is therefore clear that each HTML page
contains a considerable number of external resources, although it
should be noted that this not only includes links to script and
images, but also \textit{redirects} from a starting to a landing
webpage. Independently of the protocol and the presence of the
``www.'' prefix, the 6th column in the Table shows that redirects are
extremely popular. On average we pass via 2.4 intermediate URLs before
reaching a landing page. For this estimate we \textit{only} use
redirects that happen when loading a ``starting URL'' in our browser,
and we excluded any redirects triggered by the external resources
embedded in the HTML of the landing webpage. In general, redirects
seem to be more popular for the HTTP protocol, but the average
difference with HTTPS is minimal.

Related to the redirect phenomenon, we also notice that more than 60\%
of the crawls observed a ``change of the domain'' name among the
``starting URL'' and the ``landing URL''. This happens with only 1/3
of that frequency value if we crawl domains without the ``www.''
prefix. The reason is that often the redirection happens from one
domain to the same domain expanded with the ``www.''  prefix. A
similar trend is observable for the ``change of the protocol'', when
crawling URLs with HTTP (which get upgraded to HTTPS).

The overall number of unique IPs of the landing pages is slightly less
than 10,000, and it reflects the fact that crawling the same domain
with the four different formats, most of the time will lead to the
same landing URL/domain. On average there are 1.22 domain per each
landing IP (comparison of columns 7 and 12 in
Table~\ref{tbl:data_collection}). This is explained by the presence of
large hosting providers with many different customers. The same
argument explains why we observe a similar relationship of 1.43
domains per each IP, when considering the dataset of all
URLs. Finally, the results of \tool indicate that 89\% of the landing
URLs are served from a third-party hosting infrastructure which does
not belong to the owner of the webpage. In the following section we
illustrate how we tested the accuracy of our classification, by using
a manually validated groundtruth and by comparing with similar
applications.

\subsection{Classification Validation}
We next validate the efficacy of our tool by compiling a groundtruth 
classification, and comparing it against \tool. 

\paragraph{Compiling a Comparative Dataset} To the best of our
knowledge, no groundtruth dataset exists regarding Web hosting.  To
build this, we randomly select 324 domain names to manually annotate.
These are taken from the \textsc{top-10k} dataset, crawled with the
HTTP protocol. For each of these domains, we load the landing webpage
in a browser and use search engines to check if the owner of the IP
prefix is an organization offering Web hosting or CDN services to its
customers.

We note that 324 domains are not enough to evaluate our tool.  Thus,
we also collect equivalent data from a variety of public tools that
allow users to ``\textit{discover who is hosting a website}''. This
allows us to compare our results against their outputs.  Example of
those services include \textsc{HostingCompass.com} which can detect
who is hosting an ESLD, or \textsc{HostingDetector.com} and
\textsc{What's My CDN?} which allow more fine grained queries
including ``www.'' as prefix to the ESLD
\cite{hostingcompass,hostingdetector,whatsmycdn}. We choose to use
those three application because they are free Web-based services that
do not require any registration.  We query those services with the
URLs from our \textsc{top-10k} and \textsc{top-20k-www} datasets,
depending on the service. As the services mentioned above do not
provide any detail about the methodology they use to detect hosting
providers, in addition to those Web applications, we also use
\textsc{cdnfinder}, an open-source project which aims to detect the
usage of CDNs within websites. The tool uses \textsc{phantomjs} and a
hard-coded list of hostnames to load a webpage and detect the presence
of external resources which are hosted on a CDN~\cite{cdnfinder,
  phantomjs}. Analogously to the Web services, we downloaded the tool
and run it on our \textsc{top-40k-URLs} dataset. In total, this
results in 5 datasets to compare \tool against.

\begin{table}[t]
  \footnotesize
  \centering
  \begin{tabular}{|l|r|r|r|r|r|}
    \cline{2-5}
    \multicolumn{1}{c|}{} & \multicolumn{2}{c|}{\sc Self-hosting} & \multicolumn{2}{c|}{\sc 3rd-party hosting} & \multicolumn{1}{c}{}\\
    \hline
    {\sc Tool/Service} & {\sc TP} & {\sc FN} & {\sc TP} & {\sc FN} & {\sc F1-score}\\
    \hline
    \hline
    \tool & 26 & 3 & 279 & 16 & 0.73\\
    \hline
    hostingcompass & 27 & 2 & 102 & 193 & 0.21\\
    \hline
    hostingdetector & 12 & 17 & 239 & 56 & 0.25\\
    \hline
    whatsmycdn & 27 & 2 & 127 & 168 & 0.24\\
    \hline
    cdnfinder & 28 & 1 & 65 & 230 & 0.2\\
    \hline
  \end{tabular}
  \caption{Performance comparison of \tool and other applications on
    our manually validated groundtruth.}
  \label{tbl:manual_validation}
\end{table}

\begin{table*}[!htbp]
  \scriptsize
  \centering
  \begin{tabular}{|l|r|r|r|r|r|r|}
    \hline
    {\sc service} & {\sc domain} & {\sc www. + domain} & {\sc http + domain} & {\sc https + domain} & {\sc http + www. + domain} & {\sc https + www. + domain}\\
    \hline
    \hline
    {\sc hostingcompass} & 3,481 (3,283) & {-} & {-} & {-} & {-} & {-} \\
    \hline
    {\sc hostingdetector} & {3,229 (2,879) } & {5,607 (4,973)} & {-} & {-} & {-} & {-} \\
    \hline
    {\sc whatsmycdn} & {978 (963)} & {3,268 (3,202)} & {-} & {-} & {-} & {-} \\
    \hline
    {\sc cdnfinder} & {-} & {-} & {59 (55)} & {403 (395)} & {149 (144)} & {1,849 (1,700)} \\
    \hline
  \end{tabular}
  \caption{Comparison of \tool with similar services/applications when
    evaluated on all of our datasets. The values in the brackets
    indicates results obtained by \tool.}
  \label{tbl:comparison_with_others_on_all_datasets}
\end{table*}

\paragraph{Comparison with Manual Annotations}
Table~\ref{tbl:manual_validation} contains the results of comparing
\tool against the above online services, using the manual annotations
as the groundtruth.  Our algorithm was specifically designed for
detecting the presence of ``self-hosting'' environments. Consequently,
a domain will be flagged as ``hosted on third-parties'' in any
situation where the webpage owner differs from the owner of the
network prefix (e.g., a private Web server run at at home, where the
broadband ISP is the owner of the network prefix). Despite this
limitation, in the binary classification problem where a website is
either self-hosted or hosted on a third-party service, \tool still
outperforms all the other services that we tested, achieving an
F1-score which is almost three times larger than the average for the
other services. Indeed, on our manual groundtruth we observe an
accuracy of over 95\%, even when, instead of verifying self-hosting,
we focus on the complementary problem of detecting the presence of
third-party hosting providers. \textsc{cdnfinder} performs well in
detecting self hosting, but has a very high false positive rate when
classifying domains as third-party hosting. Similarly,
\textsc{hostingdetector} achieves the highest accuracy in detecting
external hosting services, at a cost of an extremely high false
negative rate (59\%), when a single organization is in control of both
the webpage and the network prefix.

\paragraph{Comparison with Similar Services} To further test the
accuracy of our framework, we compare our results with the four
tools/services mentioned earlier. The results of this comparison are
shown in Table~\ref{tbl:comparison_with_others_on_all_datasets}. The
goal of this is to show that \tool achieves similar results to other
applications. To this end, we narrow our goal to identify all domains
which are hosted on third-party network infrastructures. As mentioned
in the previous sections, \tool follows any kind of redirect. Hence,
Table~\ref{tbl:data_collection} only presents the classification using
the ``landing URLs/domains''. Since we do not know what are the exact
capabilities of the four services that we tested, and how they handle
redirects, we decided to back-propagate the results of our
classification from the ``landing URL'' to the corresponding
``starting URL'', and ``starting domain'', from where the navigation
started. In this way we are able to compare our results with each one
of those services and verify that our framework has a detection rate
close to those of the other services.

For almost all of the domains inspected, \tool achieves an accuracy of
around 90\%, and it identifies a third-party-hoster every time one of
the other four services detects its presence. Since the highest number
of misclassified services originate from the sets of domains analyzed
with \textsc{hostingdetector}, we sampled 20 domains without the
``www.'' prefix and another 20 with the ``www.''  prefix. We then
manually verify if they are actually hosted on a third-party
infrastructure. For 27 out of 40 cases, \textsc{hostingdetector}
failed to identify self-hosted domains and \tool correctly labeled
those as ``self-hosting''. Ten of those cases were domains of large
universities with their own network prefixes. For another 7 cases, the
landing page is the home page of large hosting services such as
``Google'', ``Salesforce'' or ``1and1''. \tool correctly labeled these
as self-hosted. For four domains our \tool did not succeed in
downloading the RDAP information, and \tool could not classify those
domains. The remaining 9 domains were hosted on a third-party
infrastructure, but we did not detect them. According to those
results, we conclude that the accuracy of \tool is inline with similar
services we compared to.


\section{Related Work}
\label{sec:relatedWork}
A significant amount of research has been done in the field of Content
Delivery Networks and cloud computing. Krishnamurthy et
al.~\cite{krishnamurthy2001use} were the first to analyze the rise of
CDNs and the benefits that they provide to end-users. After them
several studies investigated this
trend~\cite{huang2008measuring,calder2015analyzing,su2009drafting,ager2011web}. Similar
work has tried to uncover cloud usage patterns and which Web services
are running on a cloud-associated IP address~\cite{he2013next,whowas}.

Our techniques relies on a mix of methodologies, particularly
exploiting RDAP data. There have been a small set of past papers that
rely on similar data. For example, Cai et al.\ proposed to combine
WHOIS information with the ASN in order to generate a comprehensive
AS-to-organization mapping~\cite{cai2010towards}. Tajalizadehkhoob et
al.\ were the first ones to explore the identification of hosting
provides by combining passive DNS with WHOIS
information~\cite{tajalizadehkhoob2016apples}. Unfortunately their
approach leverages a classification of 2,000 ASes to filter out
organization such as ISPs, education and government. This list has
limited size and is manually generated, and this raises concerns about
its reliability across time. Contrary to previous studies, our work
does not use any precomplied list of organization names and it focuses
on identifying self-hosting environments. \tool allows other
researchers to reproduce our results and it does not require any
manual analysis or a priori knowledge of network prefixes or the ASes
in charge for routing the network traffic.


\section{Conclusion}
\label{sec:conclusion}
In this work we presented \tool, a tool for collecting information
about a webpage and the environment where the page is hosted.  Our
framework extracts information from the retrieved HTML, the DNS and
the ownership information associated to a network prefix. \tool then
exploits this data to infer if the website is self-hosted, or is
reliant on a third party operator, e.g., a Content Delivery Network.
We tested \tool on 40,000 URLs and compared the results with similar
applications that detect the presence of known hosting providers. Our
framework is accurate and outperforms all other applications, when
tested on a manually validated groundtruth. \tool is released as open
source and is built in a modular way, which gives the possibility to
integrate it with new capabilities and extensions.


\section{Acknowledgements}
\label{sec:acknowledgments}
This research was supported by the PETRAS IoT hub (EPSRC grant EP/N023242/1)
All opinions, findings and conclusions, or recommendations expressed in this
material are those of the and do not necessarily reflect the views of the
sponsors.


{
\bibliographystyle{abbrv}
\bibliography{paper}
}

\end{document}